# Using System Hyper Pipelining (SHP) to Improve the Performance of a Coarse-Grained Reconfigurable Architecture (CGRA) Mapped on an FPGA

Tobias Strauch
R&D, EDAptix, Munich, Germany
tobias@edaptix.com

*Abstract*— The well known method C-Slow Retiming (CSR) can be used to automatically convert a given CPU into a multithreaded CPU with independent threads. These CPUs are then called streaming or barrel processors. System Hyper Pipelining (SHP) adds a new flexibility on top of CSR by allowing a dynamic number of threads to be executed and by enabling the threads to be stalled, bypassed and reordered. SHP is now applied on the programming elements (PE) of a coarse-grained reconfigurable architecture (CGRA). By using SHP, more performance can be achieved per PE. Fork-Join operations can be implemented on a PE using the flexibility provided by SHP to dynamically adjust the number of threads per PE. Multiple threads can share the same data locally, which greatly reduces the data traffic load on the CGRA's routing structure. The paper shows the results of a CGRA using SHP-ed RISC-V cores as PEs implemented on a FPGA.

I. INTRODUCTION

It takes a certain time to execute a CPU instruction. Pipelining is used to improve the execution speed of a program on a single CPU. Instruction dependencies are handled by using stall signals. C-Slow Retiming (CSR) uses pipelining to **multiply** the functionality of a CPU, automatically generating a multithreaded CPU. This is a fundamentally different outcome compared to what is known when designs are pipelined. CSR is known since the 60's and outlined by Leiserson et al. in [1]. System Hyper Pipelining (SHP) improves CSR to enable more threads to be dynamically scaled on a multithreaded CPU and fits perfectly on FPGA technologies. SHP was first introduced by Strauch in [2].

At least two common problems for Multi-Processor System-On-Chips (MPSoC), Network-on-Chips (NoCs) and coarse-grained reconfigurable architectures (CGRA) can be identified. These are software (SW) partitioning challenges and potential data routing bottlenecks. In [3], Galanis et al. show the challenges to partition critical software parts on CGRAs. Various speedups can be achieved when the right method is applied. Yongjoo et al. propose in [4] memory-aware application mapping to improve the data throughput on CGRAs. The memory bandwidth optimization is discussed in [5] by Peng et al. Performance problems become even more critical when systems are mapped on FPGAs and applications need to be executed at a certain speed.

In this paper a Hyper Pipelined Reconfigurable Architecture (HPRA) is proposed, which demos improvements of the two aforementioned problems based on a CGRA. The key technology is System Hyper Pipelining, which generates multithreaded programming elements (PE) while improving the performance per area factor at the same time. This enables for example a higher local peak performance and fork-join operations, which simplifies the software partitioning problem. It also offers local data sharing, which reduces the risk of generating data routing bottlenecks. The proposed HPRA system is compared to a CGRA which uses the same routing and PEs as the HPRA, but does not use SHP to improve the performance of the PEs. The results show how a regular processor array can benefit from using SHP. It is easy to understand, how MPSoCs and NoCs can benefit from this approach as well.

SHP is outlined in Section 1. Related work to this paper is discussed in Section 2 before the novel SHP based architecture is introduced in Section 3. Results are given in Section 4.

II. CSR AND SHP TECHNOLOGY

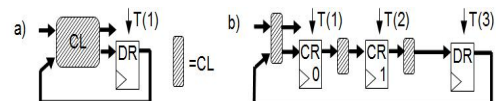

Figure 1: a) Simplified single clock design. b) Applying CSR technique.

System Hyper Pipelining (SHP) has been introduced by Strauch in [2]. This paper gives a 2-page introduction for the readers' convenience again. SHP is based on C-Slow Retiming (CSR). It enhances CSR with thread stalling, bypassing and reordering techniques by replacing the original registers of the design with memories and by adding a thread controller (TC). In the remainder of this paper, the word "thread" (T) is used synonym for the execution of a program or algorithm.

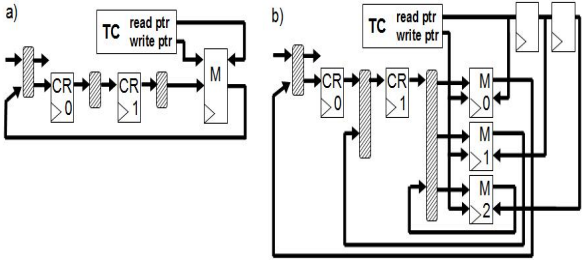

Figure 2: a) SHP-ed design with thread controller, memories and CRs. b) Further improved SHP.

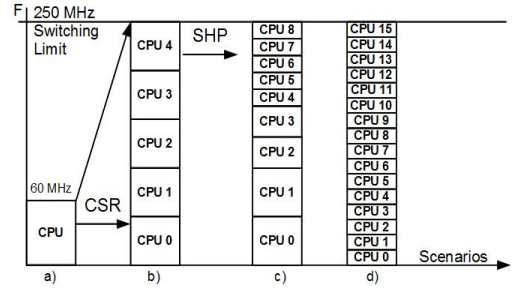

Figure 3: Histogram of different scenarios (a-d) of running CSR and SHP.

Figure 1a shows the basic structure of a sequential circuit with its inputs, outputs, combinatorial logic (CL) and original design registers (DR). The sequential circuit handles one thread T(1). Figure 1b shows the CSR technique. The original logic is sliced into C (here C=3) sections. This results in C functionally independent design copies T(C=1..3) which use the logic in a time sliced fashion. Each thread has its own thread index. For each design copy it now takes C "micro-cycles" to achieve the same result as in one cycle (called "macro-cycle") of the original design. The implemented registers are called "CSR Registers", (CR) and are placed at different C-levels (CRn).

Figure 2a shows the modifications of a CSR-ed design towards SHP. Assuming the DRs are now replaced by a memory (M). The incoming design states / threads are stored at the relevant address (write pointer) based on the thread index. D is the number of threads which the memory can hold (memory depth). The outgoing thread can now be freely selected within D available threads (read pointer), except the threads already passing through the design logic. A CSR-ed design has usually many shift registers. DRs are followed by a series of CR registers. In the SHP-ed version, many memory data outputs are connected to CRs. In this case, the shift registers at the outputs can be replaced by registers at the read address inputs of the memories (Figure 2b). The memory is sliced into individual sections (M0, M1, M2) and each section has a delayed read of the thread. The outputs can now be directly connected to the relevant combinatorial logic and the shift registers can be removed. The same trick can be applied on the shift register chains at the inputs of the memory.

$$F_{csr} = F_{orig} * C * r^C \quad (1)$$

$$0 \text{ Hz} \le F_t \le F_{orig} * r^C \quad (2)$$

$$F_{shp} = \Sigma F_t \le F_{csr} \quad (3)$$

We define $F_{orig}$ as the maximal clock rate of the original design. The maximal speed of a CSR-ed design can be estimated by using equation 1. $F_{csr}$ is C times the original speed $F_{orig}$ reduced by a correction factor $r^C$, which considers the delay inserted on the critical path by the CRs. r is technology dependent. Based on empirical data, r is roughly 0.93 for a Virtex-6 FPGA and standard designs. In an SHP-ed design, a single thread can now run at any speed (over a long period) between 0 Hz (stalled) and $F_{orig} * r^C$ (Equation 2). The maximal speed of a SHP-ed design $F_{shp}$ is the sum of all active threads (Equation 3). $F_{shp}$ cannot be greater than $F_{csr}$.

Figure 3 shows the advantages of CSR and SHP over the original design. The x-axis shows different scenarios. Assuming a single CPU runs at 60MHz on an FPGA (Figure 3a). It can be seen, how CSR improves the system performance of the original system implementation, (Figure 3b). When using CSR, the system performance is not necessarily limited by the critical path of the original design, but - for instance - by the switching limit of the FPGA (e.g. 250MHz) or the external memory access instead.

There are two key observations when SHP is used on a design. First, for executing multiple programs on multiple CPUs (symmetrical multi-processing (SMP)) or for executing multiple threads on a CPU (simultaneous multi-threading (SMT)), SHP allows a more efficient usage of the system resources. It adds the possibility to distribute the system performance over a minimum (C, Figure 3b), and a maximum (D, Figure 3c) set of design copies, whereas any solution in-between can be realized (Figure 3c). This load balancing is handled by a thread controller (TC).

Secondly, threads don't interact with each other. There is no register dependency between the individual threads. The runtime of each thread is therefore deterministic. The variable latency that the execution per thread may experience due to different behavior in if-branches for instance is not an issue, because all threads work independent of each other.

III. BASIC INTRODUCTION OF THE NOVEL ARCHITECTURE AND RELATED WORK

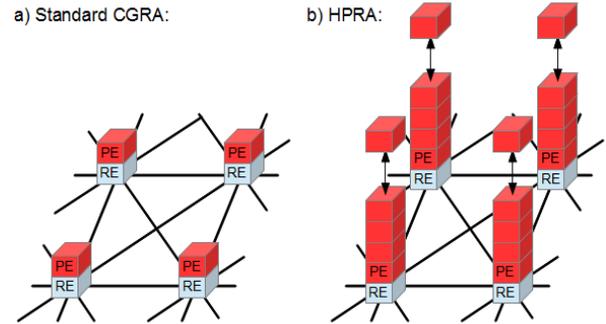

Figure 4: a) Standard CGRA with programming (PE) and routing elements (RE), b) High Performance Reconfigurable Structure with same REs but SHP-ed PEs.

Before aspects of related published work on CGRAs and NoCs are discussed, a first overview of the proposed novel architecture is given. Figure 4 gives an overview of the proposed hyper pipelined reconfigurable architecture (HPRA). The given CGRA (Figure 4a) is based on a 2 dimensional array of logic clusters, whereas each cluster has one programming element (PE) and one routing element (RE). The HPRA (Figure 4b) uses SHP to improve the performance per given area, by adding time as a 3rd dimension to generate multiple independent PE-threads which use the original PE in a time sliced fashion.

Each PE uses System Hyper Pipelining to generate multiple threads which use the logic in a time sliced fashion. One thread still runs virtually at macro-cycle (original) speed. The PE itself as well as all other elements (routing element, memory, ...) are clocked at the C times faster micro-cycle speed. Therefore the speed of the system is not necessarily dominated by the complexity of the datapath logic of the PE. The RE and the PE still run synchronously using the same clock, but it takes C micro-cycles to execute one original macro-cycle cycle of one thread.

The functional units of the application which run on the CGRA are called submodules (SM). Their individual program code is mapped to individual PEs. A PE can execute code of different SMs. Multiple threads can share the same code.

Related work must be discussed specifically with regard to the architecture's routing concept and its 3D topology. A new CGRA **routing concept** is proposed by Metzner et al. in [6]. Their proposed Quattuor-Architecture has the capability of using direct interconnects locally, within a task for fast data exchange among the submodules, and global communication using messages beyond component boundary. It tries to find an optimal trade-off between CGRA and NoC concepts. CGRAs and FPGA overlays are essentially dataflow machines to fulfill high performance requirements. A survey on CGRAs is given by Tehre et al. in [7]. Alternative concepts can be based on bus systems such as NoCs, which exchange messages among cores, memories and peripherals, all of which are connected in a network infrastructure on the chip. NoCs are usually provided as a 2-dimensional grid with routers placed at intersections between lines and columns and which are connected to homogeneous PEs. A survey on real-time NoC architectures is given by Hesham et al. in [8].

It can be said, that SHP uses time as a 3rd dimension when using the logic in a time sliced fashion. A **3D programmable logic** device has been developed by Tabula Inc. which uses operational time expansion [9] to increase performance per area. In contrast to SHP, the number of threads are fixed and individual threads cannot be stalled nor can their performance be balanced. Additionally, the logic is continuously reconfigured for each thread, which can be seen as a overhead and requires specific synthesis algorithms.

Classical **3D stacked chips** usually have problems such as that the inter-layer vias are limited in number, and the increased power density leads to high junction temperatures, as Gayasan et al. show in [10]. This interconnect bottleneck has an impact on 3D NoCs. Tradeoffs between the number of nodes utilized in the third dimension, which reduces the average number of hops traversed by a packet, and the number of physical planes used to integrate the functional blocks of the network, which decreases the length of the communication channel, is evaluated for both the latency and power consumption of a network by Pavlidis et al. in [11]. Through a detailed case study for k-ary-2-mesh networks Qian et al. have shown in [12] that transforming a 2D NoC into a 3D NoC may not improve the worst-case performance while improving the average performance.

It will be demonstrated in this paper, that SHP has a positive impact on the routing concept of a CGRA by locally sharing data, and that SHP which uses time as 3rd dimension has benefits over 3D programmable devices and stacked processor arrays.

## IV. INTRODUCTION TO THE HYPER PIPELINED RECONFIGURABLE ARCHITECTURE

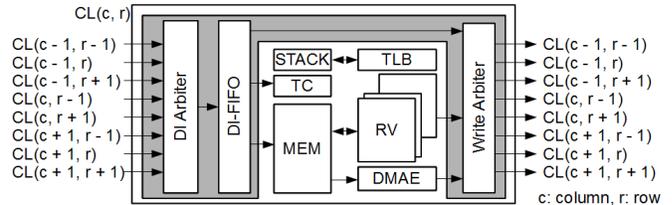

Figure 5: HPRA cluster based on routing element (gray background) and programming element.

This section discusses the hyper pipelined reconfigurable architecture (HPRA). It is based on a two dimensional array of clusters (CL), whereas each CL is based on one routing element (RE) and one programming elements (PE) as shown in Figure 5. It is also shown, how the data transfer in the system is accomplished and how the system can be partly re-configured during runtime.

### A. The RISC-V based PE

The PE is based on the RISC-V (RV) instruction-set-architecture (ISA) from Berkeley [13]. The implemented 32-bit version uses a simple RISC-V subset as well as multiply and (multi-cycle) division instructions (RISC-V32IM).

With $C = 4$ and $D = 16$ the following relative performance numbers can be estimated. If less or equal to C threads are executed, then **each** thread can run at ($r^C = 0.93^4 =$) 75% speed of the original design. If the number of active threads (T) is greater than C, then the maximal system performance, which is ($C * r^C = 4 * 75\% =$) 300% of the original design, is equally distributed over all Ts.

The system is memory limited when placed on an FPGA. Therefore certain trade-offs must be considered. The RAM of each PE (PE-RAM) is dynamically shared by instructions and data. The instructions are basically a list of (independent) program sections and functions. All Ts can execute any program section or function in that RAM and can access (read/write) the complete RAM. Data can also be written by any other PE using a mechanism which is shown later.

## B. Stack Handling

Each T needs its own stack range, which generates an immense memory overhead. Therefore all Ts share one single extra stack memory (PE-STACK) dynamically. A small register based translation-look-aside-buffer (TLB) uses stack access information as well as the current thread ID from the thread controller to enable access to a certain stack range. When the stack is full, a stack overflow is prevented by stalling the relevant threads until at least one other thread releases its section in the stack memory. The thread controller continuously executes all active threads, which automatically generates a round-robin mechanism for the stack usage when an overflow happens. This mechanism can still lead to a system stall in an extreme case. Therefore care must be taken when partitioning software on the individual PEs. The stack pointer register is set to 0 when a thread starts.

## C. The Thread Controller

Each PE has a thread controller (TC). Each TC has special-function-registers (SFR), by which the TC can be controlled. A thread (T) can be started simply by writing the T's start address to a specific address in the TC's SFR, called "Activate". The TC then assigns the T to a specific slot (S) with has a specific slot ID (SID = $\{0, ..., D - 1\}$). If more than D threads should be started, a thread-overflow occurs. Therefore care must be taken when partitioning software on the individual PEs. A handshake mechanism must be implemented on the software layer. The task runtime can vary when more than C threads are active. Multiple threads can share the same program.

A T can "kill" itself by writing (any data) to a specific SFR, called "Exit". By doing that it frees the relevant S. A T cannot be killed by other Ts. A T can also be stalled. This means that the T's design state remains in the memory M (see section 1) and is not passed though the design logic. This allows other Ts to bypass. A T can be stalled by setting the relevant bit (=SID) to a specific SFR, called "Stall". The T starts again if this bit is cleared by any other thread. Because the SID is assigned dynamically, a certain stalling mechanism must be implemented in software. Each T can read its own SID.

To enable a fork-join program execution within one PE, the following mechanism is implemented. A set of Ts an be started from a single main T (MT) by successively writing the individual start addresses of the Ts to be started to the TC's SFR called "Activate and Count (AC)". By doing that, the number of Ts called (CT) by the MT is stored in the AC register. Optionally the MT stalls itself after that process. Each CT saves the MT's SID in the "forked thread register FT". When a CT is killed, it checks the FT and decrements the AC of the MT. If this number gets 0, the MT stalling bit is cleared by default and the MT continues. Alternatively the MT can read it's AC register to continue execution.

## D. Data transfer

In the proposed architecture, each PE can run multiple threads (T). Individual Ts running on different PEs can forward data to each other using the complete HPRA's memory range. The same is true for each DMA engine, which is defined later. Data arriving at a routing element can either be forwarded to another cluster or to the PE's memory or TC. Due to the limited number of pages for this paper, this mechanism cannot be further elaborated on, but its implementation does not have a relevant impact on the achieved results shown at the end of this paper.

## E. DMA Engine

To increase the system's throughput, a direct memory access engine (DMAE) is added to each PE. The DMAE has three SFR which can be programmed by each T. The DMASA register holds the start address of the source memory and the DMAL register the transfer length. The transfer is started by writing the target address of the DMA to the DMATA register. The DMAE can only be programmed when not active.

This mechanism enables a continuous datastream throughout the system. PEs connected to a system bus can initiate a burst read on the system bus. Therefore data can be read from external using bursts.

## F. Configuration and partly reconfiguration during runtime

The system can easily be configured and partly reconfigured during runtime. For that instructions have to be streamed through the system to the target RAM by using the relevant target address. This can also be done during runtime so that parts of the applications can be reprogrammed / reconfigured. A thread is started by writing its start address to the relevant SFR of the TC.

## V. RESULTS

The proposed SHP based HPRA is now compared to the non SHP-ed CGRA using the same fast routing elements (RE) and the same DMA engine (DMAE). The programming element (PE) is based on a 3-stage RISC-V32IM (RV) as defined in [14]. Both designs are mapped on a Virtex 6 LX75T (-3ff784).

## A. Performance per Area Improvement

Table 1 compares the data of the original CGRA and the SHP-ed HPRA. The original RV occupies 617 slices (occS) and runs at 181 MHz, which results in a performance per area factor (PpA) of 0.29 MHz/occS. The SHP-ed RV (C=4, D=16) occupies 703 slices but achieves a performance of 549 MHz. The resulting PpA is 0.78 and is 266% of the original RV's PpA. Both architectures use the same DMAE so that the programming element (PE) size difference basically results from the different RV size. The routing elements (RE) are the same as well as the system support logic (SDRAM controller and system bridge). Both have a 4x4 implementation of PE/RE clusters, whereas one cluster is removed and replace by the support logic

A higher mapping effort makes it possible that each design fits into the FPGA (maximal 11640 slices). The CGRA system with the original RV implementation achieves 2.715 GHz and a PpA of 0.23 MHz/occS. The HPRA's performance is much higher (8.235 GHz) and its PpA is 203% higher (0.77 MHz/occS) than the one of the CGRA. Thus, the HPRA can execute more threads and can achieve a 3.03-times higher

Table 1. Virtex 6 based implementation of alternative concepts.

| Module | Size [occS] | Perf. [MHz] | PpA [MHz/occS] | Module | Size [occS] | Perf. [MHz] | PpA [MHz/occS] | dPpA [%] |
|---|---|---|---|---|---|---|---|---|
| RV | 617 | 181 | 0.29 | SHP-RV | 703 | 549 | 0.78 | 266 |
| PE | 697 | | | PE | 781 | | | |
| RE | 103 | | | RE | 103 | | | |
| Support | 159 | | | Support | 159 | | | |
| **CGRA** | **11634** | **2715** | **0.23** | **HPRA** | **11635** | **8235** | **0.77** | **303** |

Table 2. Local Peak Performance Using Matrix Multiplication

| | unit | 4x4 | 5x5 | 6x6 | 7x7 | 8x8 | 9x9 | 10x10 |
|---|---|---|---|---|---|---|---|---|
| RV | ns | 1039 | 1541 | 2144 | 2845 | 3646 | 4547 | 5547 |
| SHP | ns | 153 | 202 | 259 | 330 | 412 | 505 | 608 |
| Diff. | % | 670 | 762 | 829 | 863 | 886 | 901 | 912 |

Table 3. Application Partitioning Considerations

| Implementation | Threads per System | Threads per Cluster | Data Sharing per Cluster | Performance per Cluster | Performance Penalties |
|---|---|---|---|---|---|
| Single RV | 15 | 1 | no | 181 MHz | |
| SHP-ed RV min | 60 (C = 4) | 4 | yes | 549 MHz | no |
| SHP-ed RV max | 240 (D = 16) | 16 | yes | 549 MHz | Yes |

system performance than the non SHP-ed CGRA version on the same FPGA.

*B. Local Peak Performance Improvement*

In this paper, the local peak performance (LPP) defines the runtime of an algorithm based on a local data set. In other words, LPP is the execution speed of a program only accessing data available at the PE's memory and without routing data through the system. A simple matrix multiplication algorithm is used for that.

Table 2 shows the LPP for matrix multiplications of different sizes, running on the original RV implementation and a SHP-ed RV version. The SHP-ed RV can use fork-join techniques to run multiple threads in parallel. SHP allows the usage of up to D (here 16) threads. The SHP-ed RV's runtime outperforms the original version by 670% for small matrix multiplications (4x4) and 912% for larger ones (10x10).

*C. Throughput Performance*

It is assumed that the routing structure of a CGRA can be efficiently pipelined. The proposed system hyper pipelining technology is related to the programming element (here RISC-V32IM). Therefore, the routing structure of the two systems remains identical. Here a 4x4 system is used again, whereas 1 cluster is replaced by system logic (SDRAM interface and system bridge). This results in 15 clusters.

Table 3 gives an overview of what kind of aspects are important when an application needs to be partitioned over multiple threads. The standard CGRA system using single RV implementation can run up to 15 threads. Threads cannot share data on a cluster and the algorithm execution speed per thread is 181 MHz. The proposed HPRA system using SHP-ed RVs can run efficiently a minimum (min) of 60 threads (4 per cluster), whereas all threads on a cluster can share data locally. In this case, each thread on a cluster does not impact the runtime of the 3 remaining threads on a cluster, so that each thread gets its maximum share (137 MHz) of the 549 MHz cluster-performance. The system can be completely or locally scaled for running up to 260 threads on the same FPGA (max). 16 threads can be executed on a single cluster which can all share data locally. Here the performance of each thread is affected, because only 549 MHz are available per cluster. The number of threads per cluster can be adapted dynamically.

VI. CONCLUSION

C-Slow Retiming is a known technique to turn a digital design into a multithreaded version. System Hyper Pipelining (SHP) adds more flexibility to the multithreading approach. In this paper, SHP is applied on the programming element (PE) of a coarse-grained reconfigurable architecture (CGRA). One of the key advantages of the proposed system is a higher performance per area factor. The more concentric approach of running a flexible number of threads on a single PE improves system level aspects like local peak performance and throughput performance. Individual applications can overlap on PEs and can share data on the same RAM without moving them through the system.

In the proposed architecture (8*16=) 96 threads can communicate with a single PE, which itself can run up to 16 individual threads. The data throughput speed is higher compared to the execution time of a single instruction. Data routing and the dynamic memory access mechanism are independent of the PE's program execution. The various aspects discussed in this paper can easily be applied on MPSoCs and NoCs.